\documentclass[twocolumn,10pt]{article}

\usepackage[letterpaper,margin=2.25cm]{geometry}
\usepackage[T1]{fontenc}
\usepackage{newtxtext,newtxmath}
\usepackage{amsmath}
\usepackage{graphicx}
\usepackage{booktabs}
\usepackage{float}
\usepackage{enumitem}
\usepackage{cite}
\usepackage{caption}
\usepackage{url}
\usepackage[colorlinks=false,pdfborder={0 0 0}]{hyperref}
\usepackage{balance}

\newcommand{\listheading}[1]{%
 \noindent\textbf{#1}\par\nobreak\vspace{0.35\baselineskip}}

\begin{document}

\twocolumn[{%
\begin{center}
{\large\bfseries Reachability-Based Safe-Start Regions for Approach to a Tumbling Target with Rotating LOS Constraints\par}
\vskip 1.5em
{\normalsize
\textbf{Omer Burak Iskender}\textsuperscript{a,b,*},\
\textbf{Keck Voon Ling}\textsuperscript{a},\
\textbf{Wee Seng Lim}\textsuperscript{b},\
\textbf{Erick Lansard}\textsuperscript{b}\par}
\vskip 1.5em
{\normalsize\textsuperscript{a}\,\textit{Nanyang Technological University, Singapore.} E-mail: \href{mailto:iske0001@e.ntu.edu.sg}{iske0001@e.ntu.edu.sg}, \href{mailto:ekvling@ntu.edu.sg}{ekvling@ntu.edu.sg}\par}
\vskip 0.5em
{\normalsize\textsuperscript{b}\,\textit{Satellite Research Centre (SaRC), Nanyang Technological University (NTU), Singapore.} E-mail: \href{mailto:limWS@ntu.edu.sg}{limWS@ntu.edu.sg}, \href{mailto:erick.lansard@ntu.edu.sg}{erick.lansard@ntu.edu.sg}\par}
\vskip 0.5em
{\normalsize\textsuperscript{*}\,Corresponding author\par}
\vskip 1em
{\small\itshape Preprint of a paper accepted for publication at the 77th International Astronautical Congress (IAC 2026), Antalya, T\"{u}rkiye, 5--9 October 2026, paper IAC-26,C1,3,6,x110087.\par}
\end{center}
\vskip 1em
\begin{center}
\textbf{Abstract}
\end{center}
\noindent
This paper presents a closed-form test that decides, before the maneuver begins, whether a chaser can reach and hold station at the hold point of a tumbling, uncooperative target inside a line-of-sight (LOS) corridor that turns with the target. A closed-loop controller cannot answer this question: a receding-horizon controller checks the corridor only over its prediction horizon, and near a tumbling target the corridor sweeps past faster than bounded thrust can follow, so a start that is feasible at the first step can become unrecoverable later. Today the only ways to know are to fly the controller in simulation or to compute a Hamilton--Jacobi reachable set, both too slow onboard. The test combines two criteria derived from bounded-thrust relative orbital dynamics: a directional erosion margin, the corridor margin that rotation-induced drift consumes before the thruster arrests it, and a synchronization radius, beyond which the apparent rotational velocity cannot be cancelled. Guidance pairs a three-regime tracking law with a receding-horizon quadratic program. Benchmarked against polytopic backward and forward reachable sets, Hamilton--Jacobi level sets and closed-loop Monte Carlo simulation, the test runs over two orders of magnitude faster than Hamilton--Jacobi and, over 500 closed-loop cases, predicts feasibility with 91\% recall and 80\% precision. The gap to Hamilton--Jacobi is structural, not a method error: reaching the hold point and co-rotating with it is a stronger requirement than arriving with arbitrary velocity, and the gap widens with tumble rate. The test therefore gives an onboard go/no-go answer where Hamilton--Jacobi reachability is too expensive.
\par
\vskip 1.5ex
\noindent\textbf{Keywords:} proximity operations, uncooperative target, time-varying LOS corridor, safe-start region, synchronization bound, Hamilton--Jacobi reachability, MPC
\vskip 2.5ex
\noindent
\begin{minipage}[t]{0.48\textwidth}
\listheading{Nomenclature}
\noindent
$n$ \quad target orbit mean motion (rad/s) \\
$\mathbf{x}=[x,y,z,\dot{x},\dot{y},\dot{z}]^\top$ \quad 3D LVLH relative state \\
$\mathbf{u}=[a_x,a_y,a_z]^\top$ \quad LVLH control acceleration \\
$a_{\max}$ \quad maximum thrust-to-mass ratio \\
$\omega_t$ \quad target tumble rate about LVLH $z$ \\
$R_z(\theta)$ \quad rotation matrix about $z$ by angle $\theta$ \\
$c_x, c_z$ \quad LOS cone half-angle slopes \\
$x_0, z_0$ \quad docking port lateral half-widths \\
$r_h$ \quad hold stand-off radius \\
$r_{\mathrm{sync}}$ \quad synchronization range limit \\
$\Phi(\tau)$ \quad CWH state-transition matrix \\
$B_d(\tau)$ \quad CWH zero-order-hold input matrix
\end{minipage}\hfill
\begin{minipage}[t]{0.48\textwidth}
\listheading{Acronyms/Abbreviations}
\noindent
BRS: backward reachable set \\
CWH: Clohessy--Wiltshire--Hill \\
FRS: forward reachable set \\
HJ: Hamilton--Jacobi \\
IoU: intersection over union \\
LOS: line of sight \\
LQR: linear quadratic regulator \\
LVLH: local vertical local horizontal \\
MC: Monte Carlo \\
MPC: model predictive control \\
mRPI: minimal robust positively invariant \\
PD: proportional--derivative \\
PDE: partial differential equation \\
QP: quadratic program
\end{minipage}
\vskip 4.5ex
}]

\section{Introduction}
Autonomous rendezvous and docking with a tumbling uncooperative target requires guidance that respects time-varying geometric constraints imposed by the rotating docking corridor. The capability underpins on-orbit servicing and active debris removal, where the client is typically uncontrolled and slowly tumbling~\cite{Flores-Abad2014}. Tumbling is not confined to debris: a satellite typically leaves its launcher or deployer tumbling, and detumbling, the nulling of these initial angular rates, is the first phase of its early-orbit operations~\cite{ISKENDER2026112649}; a spacecraft that later loses attitude control starts tumbling again and becomes an uncooperative target of exactly the kind considered here. Classical Clohessy--Wiltshire--Hill (CWH) relative motion~\cite{Hill1878,Clohessy1960} provides a compact linear prediction model suitable for online guidance, but the constraint landscape is complicated by the target's attitude motion: the line-of-sight corridor rotates with the target body frame, producing time-varying feasibility boundaries in the chaser's coordinate system.

A critical pre-mission question is: \emph{from which initial states can the chaser safely approach and synchronize with the rotating hold point?} This paper addresses that question through a reachability-aware guidance architecture that combines:
\begin{itemize}
\item geometric safe-start region analysis via directional per-constraint erosion and a synchronization range bound,
\item a receding-horizon MPC controller~\cite{Borrelli2017,MPC2015} with CWH prediction dynamics and explicit LOS corridor constraints in the QP,
\item a three-regime tracking law that transitions from far-field LVLH approach to body-frame co-rotation and synchronized hold,
\item benchmarking of the closed-form criteria against four standard reachability engines (polytopic backward and forward reach, Hamilton--Jacobi~\cite{Mitchell2005,Bansal2017}, and closed-loop Monte Carlo).
\end{itemize}

Reachability analysis supplies the natural yardstick for such a question. Hamilton--Jacobi level-set methods characterise reachable sets exactly for low-dimensional systems and have become a standard safety tool for autonomous vehicles~\cite{Mitchell2005,ChenTomlin2018}, while set-propagation methods trade exactness for scalability~\cite{Girard2008}. The same machinery underpins adversarial formulations of non-cooperative proximity operations, where the encounter is posed as a differential game and each player's reachable set bounds what its opponent can force~\cite{iskender2026convex}; the present setting is the non-adversarial special case in which the target applies no control. A complementary line of work enforces safety through control barrier functions, which certify forward invariance of a prescribed safe set pointwise in time~\cite{Ames2019}; the criteria developed here are instead a pre-mission admissibility test on the \emph{initial} state, and are therefore closer in spirit to a reachable-set computation than to a runtime filter.

The central contribution is structural. \emph{Synchronization} (reaching the rotating hold point \emph{and} matching its co-rotation velocity $\omega_t r$, certified in closed form) and \emph{reachability} (reaching the target position with any terminal velocity, certified by the Hamilton--Jacobi backward reachable set) are not two estimates of one set but distinct objects related by the strict containment $\mathcal{S}_{\mathrm{sync}}\subsetneq\mathcal{S}_{\mathrm{reach}}$, with the gap widening as $\omega_t$ grows. Quantifying this decoupling is the operationally relevant result; the analytical synchronization bound is a sound inner certificate, not an approximation of HJ reachability.

\section{Dynamics Model}

\subsection{CWH Relative Motion}
Both the guidance prediction model and the truth propagation use the three-dimensional CWH linearised relative dynamics in the LVLH frame. The continuous-time equations are:
\begin{align}
\ddot{x} &= 3n^2 x + 2n\dot{y} + a_x, \notag \\
\ddot{y} &= -2n\dot{x} + a_y, \label{eq:cwh} \\
\ddot{z} &= -n^2 z + a_z. \notag
\end{align}

The exact discrete-time state-transition matrix $\Phi(\tau)$ and zero-order-hold input matrix $B_d(\tau)$~\cite{Schaub2018} are used for both MPC horizon predictions and truth propagation. Writing $c = \cos(n\tau)$, $s = \sin(n\tau)$:
\begin{figure*}[!t]
\begin{equation}
\Phi(\tau) = \begin{bmatrix}
4 - 3c & 0 & 0 & s/n & 2(1-c)/n & 0 \\
6(s - n\tau) & 1 & 0 & -2(1-c)/n & (4s - 3n\tau)/n & 0 \\
0 & 0 & c & 0 & 0 & s/n \\
3ns & 0 & 0 & c & 2s & 0 \\
-6n(1-c) & 0 & 0 & -2s & 4c-3 & 0 \\
0 & 0 & -ns & 0 & 0 & c
\end{bmatrix}.
\label{eq:phi}
\end{equation}
\end{figure*}

Truth propagation uses sub-stepping (3 sub-steps per control interval) with this exact matrix exponential to maintain numerical accuracy.

\subsection{Physically Honest Propagation}
A key design principle is that the truth simulation uses \emph{only} CWH dynamics plus the commanded acceleration: no reference blending, position projection, or velocity overrides are applied. This ensures that all feasibility claims are physically honest: if the chaser reports ``hold achieved'', it genuinely reached the hold band through dynamically consistent motion.

\section{Backward Reachability: From Double Integrator to CWH}
\label{sec:three_forms}

Before adding the rotating line-of-sight corridor, we build intuition on the backward reachable set itself, first on a simple double integrator and then on the orbital CWH dynamics. This isolates two effects that the constrained safe-start analysis (\S\ref{sec:safe_start}) later inherits: how process uncertainty shapes the reachable set, and how orbital coupling alone deforms it relative to a naive double integrator.

The Backward Reachable Set admits three classical variants distinguished by
their treatment of process uncertainty, a canonical sequence in standard
predictive-control references~\cite{Borrelli2017,MPC2015}:

\begin{itemize}
\item \textbf{Nominal BRS:} $x_0 \in \mathcal{X}_0^{\mathrm{nom}}$ iff
$\exists u_0, \dots, u_{N-1}$ such that $x_N \in \mathcal{T}$ under the
deterministic dynamics $x_{k+1} = A x_k + B u_k$.
\item \textbf{Stochastic BRS:} additive Gaussian noise $w_k \sim
\mathcal{N}(0, \Sigma_w)$ enters the recursion as
$x_{k+1} = A x_k + B u_k + w_k$. The BRS is computed by Monte Carlo
over the noise samples; it gives an empirical reach set that contracts
toward the nominal as $\Sigma_w \to 0$.
\item \textbf{Robust BRS:} bounded disturbance $w_k \in \mathcal{W}$ (a box);
the predecessor step is accepted only when feasible for every vertex of
$\mathcal{W}$. The resulting set is the largest sub-set of the nominal BRS
that tolerates any disturbance realization, equivalent to the
Mayne--Seron--Rakovi\'c rigid tube cross-section~\cite{Mayne2005} in
its 1-step form.
\end{itemize}

\begin{figure*}[!t]
\centering
\includegraphics[width=\textwidth]{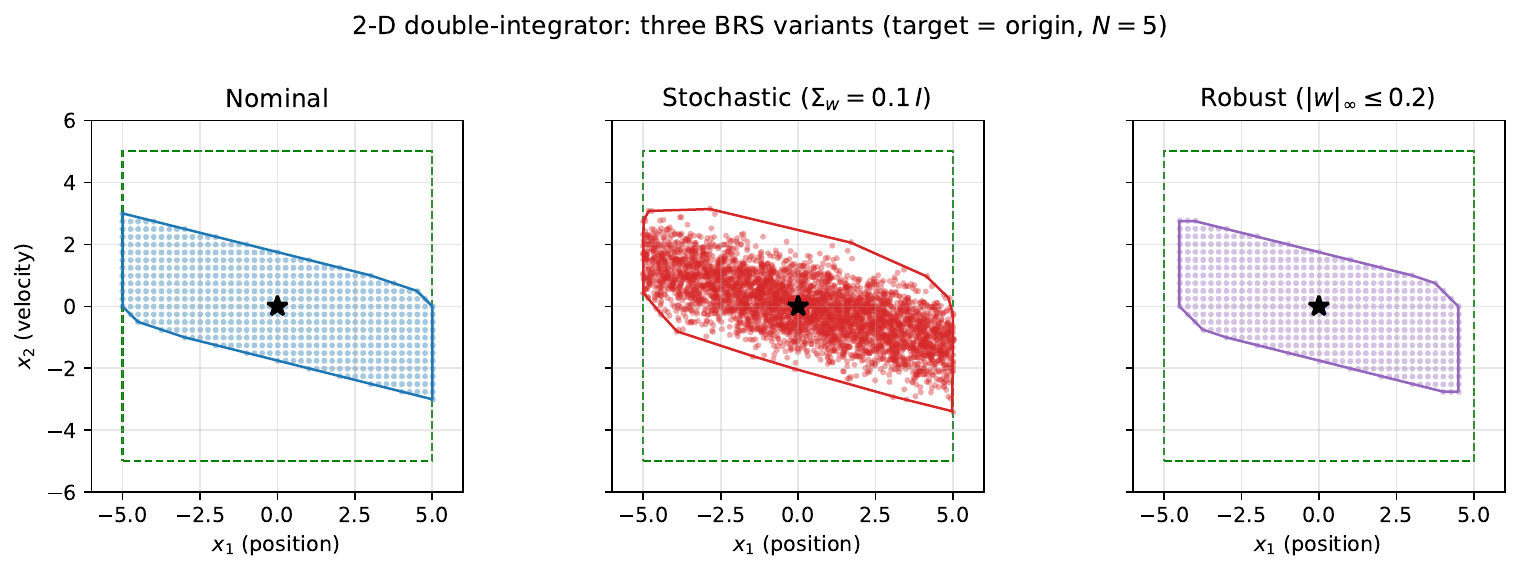}
\caption{Nominal, stochastic, and robust backward reachable sets on the
canonical double-integrator ($u \in [-1, 1]$, $|x_i| \le 5$, target = origin,
$N = 5$). Robust $\subseteq$ nominal; the stochastic cloud brackets the nominal
boundary (see text).}
\label{fig:di_three_forms}
\end{figure*}

Figure~\ref{fig:di_three_forms} illustrates these on the canonical
double-integrator $A = \left[\begin{smallmatrix}1 & 1\\ 0 & 1\end{smallmatrix}\right]$, $B = [0, 1]^\top$,
$|u| \le 1$, $|x_i| \le 5$, with target at the origin and horizon $N = 5$.
The robust set is a strict subset of the nominal set,
$\mathcal{X}_0^{\mathrm{robust}} \subseteq \mathcal{X}_0^{\mathrm{nom}}$
(it must tolerate every disturbance-box vertex), while the stochastic set is an
empirical sample that brackets the nominal boundary and expands with the noise
covariance.

\begin{figure*}[!t]
\centering
\includegraphics[width=\textwidth]{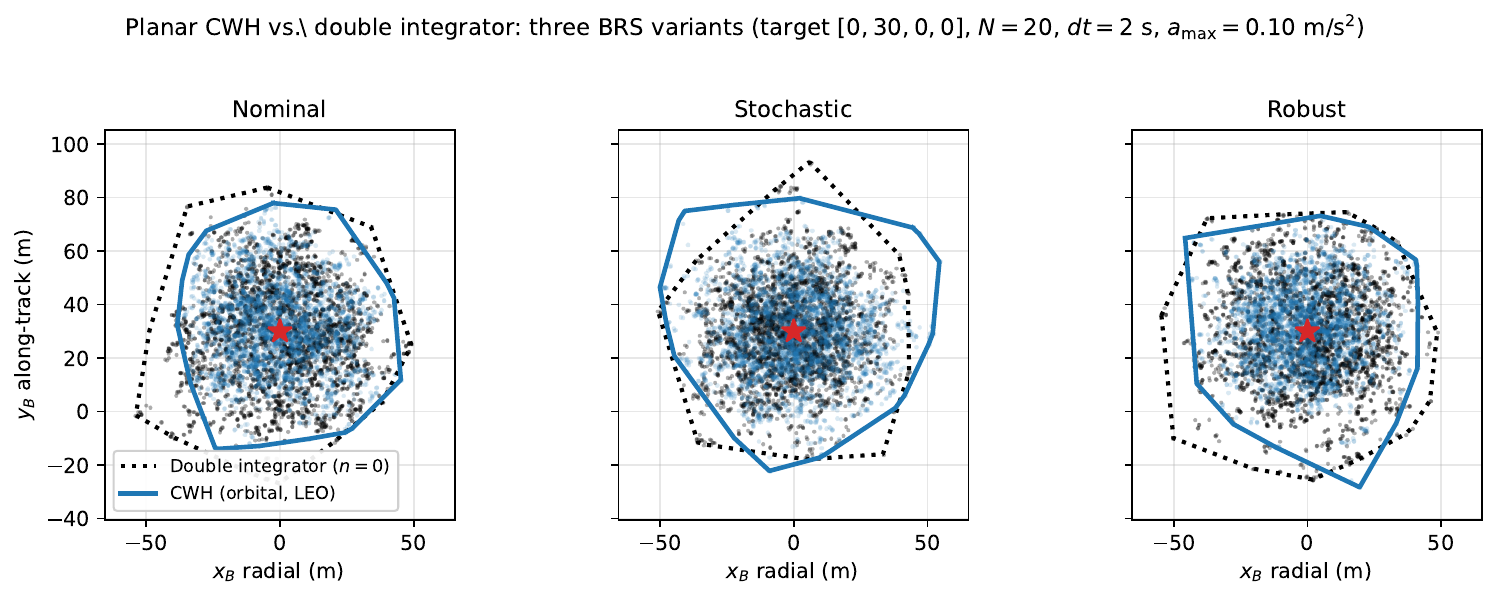}
\caption{Planar CWH ($n = 1.1\times10^{-3}$~rad/s, blue solid hull) vs.\ a
same-order double integrator ($n = 0$, grey dashed hull) backward reachable set
for the three BRS variants; target hold $[0, 30, 0, 0]^\top$, $N = 20$,
$dt = 2$~s, $a_{\max} = 0.10$~m/s$^2$. Orbital dynamics contract the radial
($x_B$) reach to $\sim$81\% of the double integrator's (see text).}
\label{fig:cwh_three_forms}
\end{figure*}

\textbf{Mapping to the present problem.}
For the CWH+rotating-LOS setting of \S\ref{sec:safe_start}, the three forms
specialize as follows.
\emph{Nominal} BRS corresponds to the deterministic LOS-tightening described
in \S\ref{sec:catalog} and benchmarked in
\S\ref{sec:benchmark}; this is the engine reported as
``Backward-Reach'' in Table~\ref{tab:benchmark}.
\emph{Stochastic} BRS models thruster jitter / navigation noise as Gaussian
process disturbance and produces an empirical sample of feasible ICs;
\emph{robust} BRS models bounded thruster calibration error and produces
a strict inner approximation of the nominal set tolerated against the
worst-case disturbance realization. Figure~\ref{fig:cwh_three_forms} contrasts
the planar CWH backward reachable set with that of a same-order double
integrator (the CWH model with mean motion $n=0$) over a realistic
$N = 20$ (40~s) terminal-approach horizon, isolating the effect of the
orbital dynamics: the only difference between the two models is the Coriolis and
gravity-gradient coupling of the CWH equations. The orbital dynamics
\emph{contract the radial} ($x_B$) reach to $\sim$81\% of the double
integrator's (nominal radial range $\approx 84$~m vs $103$~m): the coupling
curves radial excursions into along-track drift, so fewer radial offsets can be
driven back to the hold point within the horizon, while the along-track reach is
comparable; the contraction grows with the horizon-to-orbit-period ratio. In
each panel the solid hull is CWH and the dashed hull the double integrator;
across panels the robust set is a subset of the nominal set for each model, with
the stochastic cloud bracketing the nominal boundary. (Point clouds in
Figs.~\ref{fig:di_three_forms}--\ref{fig:cwh_three_forms} are uniformly
subsampled for rendering; the set extents, not the marker counts, carry the
comparison.)

The robust BRS is conceptually the closest of the three to the analytical
synchronization criterion: both require feasibility under the worst-case
realization of an exogenous perturbation (bounded disturbance for robust BRS,
worst-case rotation-induced drift for the analytical bound).
Code reproducing Figs.~\ref{fig:di_three_forms} and~\ref{fig:cwh_three_forms}
is available in the open-source repository accompanying this paper.

Having characterised the unconstrained reachable set and the deformation introduced by orbital dynamics, we now add the rotating LOS corridor, which turns the safe-start question into a time-varying constrained problem.

\section{Time-Varying LOS Corridor}\label{sec:los}

\subsection{3D Body-Frame Formulation}
Let $\mathbf{p} = [x, y, z]^\top$ denote relative position in LVLH. The target body frame rotates about the LVLH $z$-axis at rate $\omega_t$:
\begin{equation}
\mathbf{p}_B = R_z(-\omega_t t)\,\mathbf{p}.
\end{equation}

In body-frame coordinates $[x_B, y_B, z_B]^\top$, the LOS corridor is a polyhedral cone defined by five half-space inequalities; constraint tightening of this polyhedral corridor for MPC-based rendezvous and docking with uncooperative targets is developed in~\cite{iskender2018constraints,burak2020model}, and for elliptical orbits in~\cite{Iskender2026EllipticalTube,iskender2026horizon}:
\begin{equation}
\underbrace{\begin{bmatrix}
0 & -1 & 0 \\
1 & -c_x & 0 \\
-1 & -c_x & 0 \\
0 & -c_z & 1 \\
0 & -c_z & -1
\end{bmatrix}}_{A_c}
\begin{bmatrix} x_B \\ y_B \\ z_B \end{bmatrix}
\le
\underbrace{\begin{bmatrix}
-r_h \\
x_0 - c_x r_h \\
x_0 - c_x r_h \\
z_0 - c_z r_h \\
z_0 - c_z r_h
\end{bmatrix}}_{b_c}.
\label{eq:los_3d}
\end{equation}

With parameters $c_x = c_z = 1.5$, $x_0 = z_0 = 2.5$~m, and $r_h = 0.5$~m, the corridor opens at approximately $56^\circ$ half-angle in both lateral directions.

\subsection{Body--LVLH Coordinate Transforms}
The body-to-LVLH velocity transform includes the Coriolis transport term:
\begin{equation}
\mathbf{v}_L = R_z(\theta)\left(\mathbf{v}_B + \boldsymbol{\omega} \times \mathbf{r}_B\right),
\quad \boldsymbol{\omega} = [0, 0, \omega_t]^\top.
\end{equation}
This is critical for correct velocity matching: at range $r$, a body-frame--stationary chaser must maintain co-rotation velocity $v_{\mathrm{corot}} = \omega_t r$ in the LVLH frame.

\section{Guidance Architecture}\label{sec:guidance}

\subsection{Three-Regime Controller}
The controller operates in three regimes based on range $r = \|\mathbf{r}\|$. The far/close handover occurs at the body-frame tracking threshold $r_{\mathrm{track}} = a_{\max}T_{\mathrm{ramp}}/\omega_t$, with ramp budget $T_{\mathrm{ramp}} = 10\Delta t$: the range below which the co-rotation velocity $\omega_t r$ can be built up within $T_{\mathrm{ramp}}$. This is a tuning gain of the tracking law and is deliberately distinct from the analytical synchronization radius $r_{\mathrm{sync}} = 2a_{\max}/\omega_t^2$ of~\eqref{eq:rsync}, which is a braking bound on the thrust authority; the two agree only when $\omega_t T_{\mathrm{ramp}} = 2$. No result below depends on their agreement, since the certificate of \S\ref{sec:safe_start} is evaluated independently of the controller.

\begin{enumerate}
\item \textbf{Far approach} ($r > r_{\mathrm{track}}$): LVLH-frame PD tracking of the spiral reference with velocity limiting. The achievable velocity is capped at $v_{\mathrm{achv}} = \min(a_{\max} t / 2,\; v_{\max})$, preventing the controller from demanding velocities that exceed the thrust budget.

\item \textbf{Close approach} ($r \le r_{\mathrm{track}}$): body-frame PD tracking. The reference is transformed to body-frame coordinates, and the controller tracks the body-frame position and velocity errors. This regime naturally handles the co-rotation requirement.

\item \textbf{Hold} ($r_h - \epsilon \le r \le r_h + \epsilon$, $v < v_{\mathrm{switch}}$): synchronized station-keeping at the hold point. A stiffer PD law with CWH gravity-gradient feedforward maintains the chaser at the rotating hold position:
\begin{equation}
\mathbf{a}_{\mathrm{ff}} = -\begin{bmatrix} 3n^2 x + 2n\dot{y} \\ -2n\dot{x} \\ -n^2 z \end{bmatrix}.
\end{equation}
\end{enumerate}

\subsection{Optimization Problem: Cost and Constraints}
The nominal regime-tracking command $\mathbf{u}_{\mathrm{nom}}$ of the previous subsection is refined at each step $k$ by a lightweight \emph{single-input receding-horizon safety filter}. Rather than optimise a full input sequence, the filter selects the single current-step acceleration $\mathbf{u}$ that stays closest to $\mathbf{u}_{\mathrm{nom}}$ while keeping the constant-input CWH forecast inside the rotating LOS corridor over the next $N$ nodes:
\begin{equation}
\begin{aligned}
\min_{\mathbf{u}\in\mathbb{R}^3}\quad
& \|\mathbf{u}-\mathbf{u}_{\mathrm{nom}}\|^2
 + w_u\,\|\mathbf{u}\|^2
 + w_{\Delta u}\,\|\mathbf{u}-\mathbf{u}_{k-1}\|^2 \\
\text{subject to}\quad
& \hat{\mathbf{x}}_j=\Phi_j\,\mathbf{x}_k+B_{d,j}\,\mathbf{u},\ \ j=1,\dots,N, \\
& A_c\,R_z(-\theta_j)\,\hat{\mathbf{r}}_j\le b_c-\epsilon_m,\ \ j=1,\dots,N, \\
& |u_i|\le a_{\max},\ \ i=1,2,3,
\end{aligned}
\label{eq:mpc_ocp}
\end{equation}
followed by a Euclidean projection $\mathbf{u}\leftarrow a_{\max}\,\mathbf{u}/\max(\|\mathbf{u}\|_2,\,a_{\max})$ onto the thrust disk. Here $\hat{\mathbf{r}}_j$ is the position block of the constant-input forecast $\hat{\mathbf{x}}_j=\Phi_j\,\mathbf{x}_k+B_{d,j}\,\mathbf{u}$ (the \emph{same} command $\mathbf{u}$ is held across the forecast, so $\hat{\mathbf{r}}_j=\Phi_j[1{:}3,:]\,\mathbf{x}_k+B_{d,j}[1{:}3,:]\,\mathbf{u}$), $\Phi_j=\Phi(j\Delta t)$ and $B_{d,j}=B_d(j\Delta t)$ are the CWH state-transition and input matrices~\eqref{eq:phi}, $\theta_j=\omega_t(t_k+j\Delta t)$ is the corridor rotation angle that makes the LOS constraint~\eqref{eq:los_3d} time-varying, $\mathbf{u}_{k-1}$ is the previously applied command, and $\epsilon_m$ is a small constraint-tightening margin. The thrust limit is imposed as the axis-aligned box $|u_i|\le a_{\max}$ inside the (three-variable) quadratic program and then tightened to the exact Euclidean disk $\|\mathbf{u}\|_2\le a_{\max}$ by the post-solve projection, so the delivered command always respects the disk; the filter also carries single-step radial keep-out and per-axis speed-limit rows, omitted here for brevity.

The cost penalises both the control magnitude and its increment, which is standard and well-posed: the magnitude term $w_u\|\mathbf{u}\|^2$ proxies fuel ($\Delta V$) expenditure and regularises the program, while the move-suppression term $w_{\Delta u}\|\mathbf{u}-\mathbf{u}_{k-1}\|^2$ smooths the command and suppresses chattering across the regime handovers. The two are complementary rather than redundant; setting $w_{\Delta u}=0$ recovers the pure effort-penalised filter and $w_u=0$ the pure rate-penalised one. We use $w_u=0.1$ and $w_{\Delta u}=2.0$ (rate-dominated), horizon $N=6$ ($12$~s at $\Delta t=2$~s), solved by sequential quadratic programming. The filter is deliberately cheap: it inherits the regime logic from $\mathbf{u}_{\mathrm{nom}}$ and uses the horizon only to keep the held command corridor-feasible, consistent with the low-cost onboard-guidance theme of this paper. A full sequence-optimising MPC over $\mathbf{u}_{0:N-1}$ with the same prediction model and corridor constraints is a natural extension.

\subsection{CWH Gravity-Gradient Feedforward}
In both the hold and tracking regimes, the controller compensates the CWH gravity-gradient acceleration. This is essential because the radial--along-track coupling ($3n^2 x$ and $2n\dot{y}$, $-2n\dot{x}$ terms) produces secular drift in the along-track direction if left uncompensated. Without feedforward, a chaser near the target drifts away due to the $6(s - n\tau)$ term in the state-transition matrix.

\section{Safe-Start Region Analysis}\label{sec:safe_start}

The safe-start analysis determines, for each body-frame position $(x_B, y_B)$ inside the LOS cone, whether a chaser starting at rest ($v_0 = 0$) can maintain constraint feasibility. Two conservative criteria are applied.

\subsection{Directional Per-Constraint Erosion}
The rotation of the body frame at rate $\omega_t$ creates an apparent velocity for an inertially-stationary chaser. For each constraint face $i$ in~\eqref{eq:los_3d}, the margin consumed during the settling transient before the thruster arrests the drift is:
\begin{equation}
\delta_i = \frac{(\dot{s}_i^-)^2}{2\,a_{\max}},
\label{eq:erosion}
\end{equation}
where $\dot{s}_i^- = \min(0, \dot{s}_i)$ is the negative part of the constraint-slack rate. The point is declared safe if $s_i - \delta_i > 0$ for all constraints $i$.

\subsection{Synchronization Range Bound}
At range $r$, the apparent rotational speed is $v_{\mathrm{rot}} = \omega_t r$. The braking distance to cancel this velocity is $d_{\mathrm{brake}} = \omega_t^2 r^2 / (2a_{\max})$. Requiring $d_{\mathrm{brake}} < r$ gives:
\begin{equation}
r < r_{\mathrm{sync}} = \frac{2\,a_{\max}}{\omega_t^2}.
\label{eq:rsync}
\end{equation}

\subsection{Coverage Maps}
Table~\ref{tab:reachability_sweep} reports the safe fraction of the LOS cone for each tumble-rate and thrust-authority combination, combining the erosion criterion~\eqref{eq:erosion} and the synchronization range bound~\eqref{eq:rsync}.

\begin{table}[H]
\centering
\caption{Safe fraction of the body-frame LOS cone for various tumble-rate and thrust-authority combinations.  Criteria: directional per-constraint erosion $\delta_i = \frac{1}{2}\dot{s}_i^2 / a_{\max}$ and synchronization range bound $r < 2a_{\max}/\omega_t^2$.}
\label{tab:reachability_sweep}
\begin{tabular}{c|cccc}
\hline
 & \multicolumn{4}{c}{$a_{\max}$ (m/s$^2$)} \\
$\omega_t$ (deg/s) & 0.20 & 0.10 & 0.05 & 0.01 \\
\hline
1 & 84.3\% & 76.9\% & 68.7\% & 8.2\% \\
2 & 68.7\% & 47.6\% & 12.7\% & 0.5\% \\
3 & 39.7\% & 10.0\% & 2.6\% & 0.1\% \\
4 & 12.7\% & 3.2\% & 0.9\% & 0.0\% \\
5 & 5.3\% & 1.4\% & 0.4\% & 0.0\% \\
\hline
\end{tabular}
\end{table}

\begin{figure*}[!t]
\centering
\includegraphics[width=\textwidth]{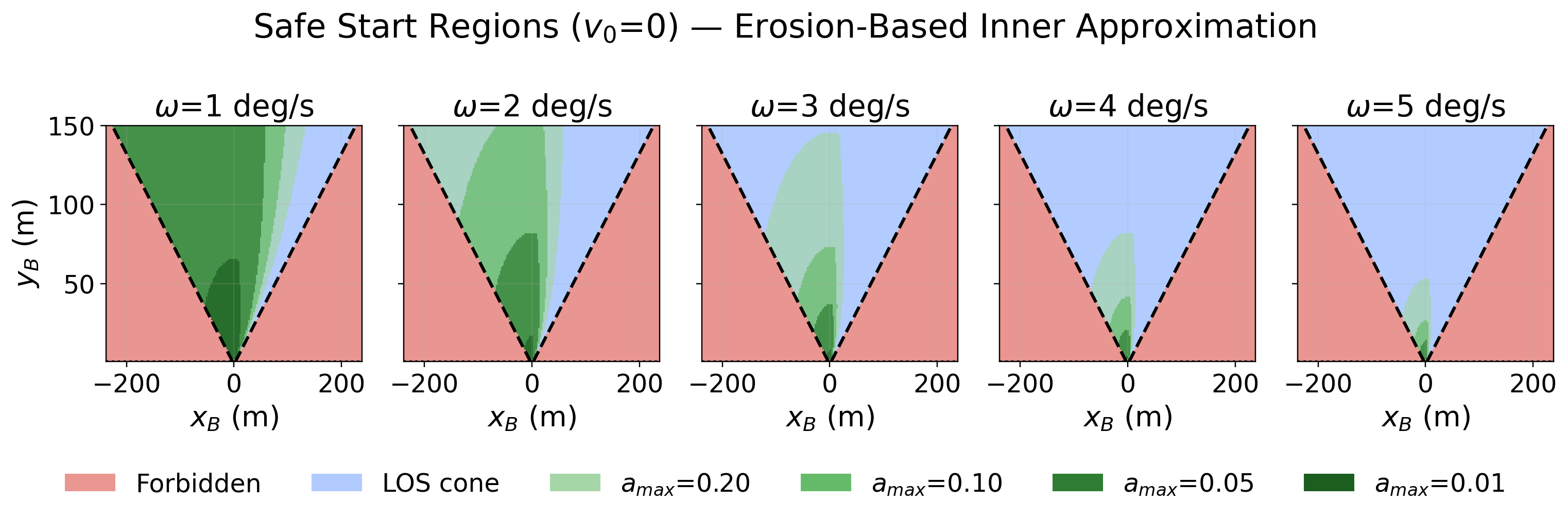}
\caption{Combined safe-start region overview for five tumble rates ($\omega_t = 1, 2, 3, 4, 5$~deg/s). Each panel shows the LOS cone (blue) and forbidden exterior (red), with nested safe-start regions for four thrust-authority levels $a_{\max}\in\{0.20,0.10,0.05,0.01\}$~m/s$^2$ at $v_0=0$. The safe region fills most of the cone at low tumble rate and collapses toward the hold point as $\omega_t$ grows ($\propto\omega_t^{-2}$ via $r_{\mathrm{sync}}$).}
\label{fig:reach_overview}
\end{figure*}

Fig.~\ref{fig:reach_overview} shows the combined body-frame safe-start maps for five tumble rates ($\omega_t = 1$--$5$~deg/s). The safe region shrinks both due to per-constraint erosion (asymmetric narrowing) and the synchronization range bound (circular cutoff at $r_{\mathrm{sync}}$).

\section{Reachability Methods Catalog}\label{sec:catalog}

To validate the analytical criteria of~\S\ref{sec:safe_start}, four standard reachability methods are applied to the same CWH + rotating-LOS problem:

\paragraph{Backward Reachable Set (BRS).} The set of initial conditions from which the rotating LOS target is reachable in $N$ steps. Rather than enumerate the $2^{2N}$ vertices of the input box (which caps the horizon at $N\le7$), we test the $v_0=0$ slice directly by linear-program feasibility: a cell is in the BRS iff some admissible input sequence $u_{0:N-1}\in[-a_{\max},a_{\max}]^{2N}$ drives the CWH-predicted position into the rotating LOS cone at step $N$. This is tractable at any horizon and is evaluated at the same horizon as HJ ($N=20$, 40~s); at that matched horizon BRS agrees with HJ to IoU $\approx0.76$ (both engines use the per-axis input bound $\|u\|_\infty\le a_{\max}$; the residual is numerical, discrete-time LP reachability vs the continuous-time PDE solver's dissipation).

\paragraph{Forward Reachable Set (FRS).} Support-function propagation~\cite{Girard2008} on $n_{\rm dirs} = 48$ template directions, with recursion
\begin{equation}
h_{R_{k+1}}(d) = h_{R_k}(A^\top d) + a_{\max}\,\|B^\top d\|_1.
\end{equation}
An initial condition is declared feasible iff at least one vertex of the target polygon lies inside the outer-polytope FRS at step~$N$, a necessary condition for true reachability, yielding an over-approximation of the safe set with known false-positive risk.

\paragraph{Hamilton--Jacobi (HJ).} Level-set solution of the HJ PDE
\begin{equation}
\frac{\partial V}{\partial t} + \min_u \nabla V^\top (Ax + Bu) = 0,
\end{equation}
on the planar~4D slice $(x_B, \dot{x}_B, y_B, \dot{y}_B)$ with grid $31{\times}31{\times}15{\times}15$ using the JAX-based \texttt{hj\_reachability} toolbox~\cite{HJReachability}, which provides the ground-truth backward reachable set up to grid resolution. The target is the LOS cone at the terminal instant $\theta_N=\omega_t N\Delta t$ and no running minimum is taken over the horizon, so this is a terminal-time reach set, not a reach tube.

\paragraph{Monte Carlo (MC).} Closed-loop rollout of the MPC controller of~\S\ref{sec:guidance} from each grid initial condition, labelled safe iff no LOS violation occurs in $T_{\rm sim}$. Empirical baseline closest to operational reality.

Related tube-MPC work~\cite{Mayne2005,Iskender2026EllipticalTube,iskender2026horizon,Iskender2026Sloshing}, including fuel-slosh--aware constraint tightening, addresses disturbance-rejection problems orthogonal to our time-varying-constraint setting; Lyapunov-funnel region-of-attraction estimation~\cite{ReistTedrake2010,Maywald2021,Shala2024} addresses post-capture trajectory-tracking certification and is discussed in~\S\ref{sec:roa}.

\section{Benchmark Results}\label{sec:benchmark}

\subsection{Synchronization vs Reachability}
At the representative operating point $\omega_t = 3$~deg/s, $a_{\max} = 0.10$~m/s$^2$ ($v_0 = 0$; grid $31{\times}31$, or $21{\times}21$ for the far costlier closed-loop Monte Carlo), the five safe-start engines reveal a central finding: the analytical region is a strict subset of the HJ reachable set, with IoU~$\approx 0.07$ (Table~\ref{tab:benchmark}). The gap is not a method error: the analytical criteria measure \emph{synchronization} (the chaser must both reach and co-rotate at the hold-point velocity $v_{\mathrm{corot}} = \omega_t r$), whereas the HJ reach set measures \emph{reachability} (the chaser can reach the target position at terminal time with any velocity). The synchronization requirement is strictly stronger; the analytical region is the operationally relevant subset for missions that require sustained co-rotation. All engine regions are restricted to the LOS corridor, since an initial state outside the cone violates the corridor at $t=0$ and cannot be a safe start regardless of terminal reachability; within the corridor the reachability engines (BRS/FRS/HJ) fill most of the cone while the synchronization-aware analytical and closed-loop MPC regions are markedly smaller.

Table~\ref{tab:benchmark} reports IoU vs HJ, area ratio, compute time, and speedup for each method at three operating points.

\begin{table*}[!t]
\centering
\caption{Benchmark comparison across three operating points ($a_{\max} = 0.10$~m/s$^2$). The analytical region shrinks relative to HJ as $\omega_t$ grows ($r_{\mathrm{sync}}\propto\omega_t^{-2}$), the cost of insisting on synchronization.}
\label{tab:benchmark}
\begin{tabular}{llcccc}
\toprule
$(\omega_t~[\text{deg/s}],\, a_{\max}~[\text{m/s}^2])$ & Method & IoU vs HJ & area / area$_{\rm HJ}$ & $t$ (s) & speedup vs HJ \\
\midrule
(1, 0.10) & Analytical & 0.375 & 0.508 & 0.01 & 177.2$\times$ \\
(1, 0.10) & Backward-Reach & 0.762 & 1.313 & 1.57 & 1.5$\times$ \\
(1, 0.10) & Forward-Reach & 0.831 & 1.065 & 65.99 & 0.036$\times$ \\
(1, 0.10) & HJ & 1.000 & 1.000 & 2.40 & 1.0$\times$ \\
(1, 0.10) & MC & 0.372 & 0.518 & 751.69 & 0.0032$\times$ \\
\midrule
(3, 0.10) & Analytical & 0.073 & 0.073 & 0.01 & 264.9$\times$ \\
(3, 0.10) & Backward-Reach & 0.762 & 1.312 & 1.82 & 1.3$\times$ \\
(3, 0.10) & Forward-Reach & 0.811 & 1.080 & 66.12 & 0.036$\times$ \\
(3, 0.10) & HJ & 1.000 & 1.000 & 2.37 & 1.0$\times$ \\
(3, 0.10) & MC & 0.045 & 0.049 & 849.10 & 0.0028$\times$ \\
\midrule
(5, 0.10) & Analytical & 0.008 & 0.008 & 0.01 & 287.9$\times$ \\
(5, 0.10) & Backward-Reach & 0.749 & 1.335 & 1.61 & 1.4$\times$ \\
(5, 0.10) & Forward-Reach & 0.809 & 1.115 & 65.90 & 0.035$\times$ \\
(5, 0.10) & HJ & 1.000 & 1.000 & 2.30 & 1.0$\times$ \\
(5, 0.10) & MC & 0.004 & 0.004 & 629.38 & 0.0037$\times$ \\
\bottomrule
\end{tabular}
\end{table*}

The analytical engine completes in $\sim 10$~ms vs HJ's $\sim 2.4$~s (a $\sim 250\times$ speedup), making it suitable for onboard go/no-go decisions where HJ is prohibitively expensive.

\subsection{Closed-Loop Sweep: Certified vs Empirical Feasibility}
Table~\ref{tab:sweep500} reports the outcome of a 500-case sweep over $a_{\max} \in \{0.02, 0.05, 0.10\}$~m/s$^2$, $\omega_t \in \{1, 2, 3, 4, 5\}$~deg/s, $r_0 \in \{30, 50, 100, 150\}$~m, with 9 randomized initial conditions per $(a_{\max}, \omega_t, r_0)$ cell; the $a_{\max}=0.10$~m/s$^2$, $\omega_t=5$~deg/s row is omitted and one cell holds 5 cases (see the table notes), giving 500 cases in total. Each case is labelled \emph{certified} (inside the analytical safe set) and \emph{feasible} (closed-loop MPC achieves $T_{\rm sim} = 60$~s with zero LOS violations), yielding three feasibility regions: certified-safe (both true), empirical-only (feasible but not certified), and unsafe (closed-loop violated).

\begin{table*}[!t]
\centering
\caption{500-case sweep results: per $(a_{\max}, \omega_t)$ row, counts of \emph{certified}/\emph{feasible} initial conditions across 9 randomized starts at each $r_0$ (m). Each cell totals 9 unless otherwise noted.}
\label{tab:sweep500}
\begin{tabular}{cc cccc}
\toprule
 & & \multicolumn{4}{c}{$r_0$ (m): certified / feasible (out of 9)} \\
\cmidrule(lr){3-6}
$a_{\max}$ (m/s$^2$) & $\omega_t$ (deg/s) & 30 & 50 & 100 & 150 \\
\midrule
0.02 & 1 & 7 / 7 & 8 / 8 & 5 / 6 & 0 / 8 \\
0.02 & 2 & 7 / 3 & 0 / 0 & 0 / 0 & 0 / 0 \\
0.02 & 3 & 0 / 0 & 0 / 0 & 0 / 0 & 0 / 0 \\
0.02 & 4 & 0 / 0 & 0 / 0 & 0 / 0 & 0 / 0 \\
0.02 & 5 & 0 / 0 & 0 / 0 & 0 / 0 & 0 / 0 \\
\midrule
0.05 & 1 & 9 / 9 & 9 / 9 & 7 / 7 & 7 / 8 \\
0.05 & 2 & 6 / 6 & 6 / 5 & 0 / 0 & 0 / 0 \\
0.05 & 3 & 5 / 1 & 0 / 0 & 0 / 0 & 0 / 0 \\
0.05 & 4 & 0 / 0 & 0 / 0 & 0 / 0 & 0 / 0 \\
0.05 & 5 & 0 / 0 & 0 / 0 & 0 / 0 & 0 / 0 \\
\midrule
0.10 & 1 & 9 / 9 & 9 / 5 & 5 / 4 & 9 / 9 \\
0.10 & 2 & 9 / 9 & 6 / 8 & 5 / 4 & 4 / 1 \\
0.10 & 3 & 6 / 6 & 7 / 2 & 0 / 0 & 0 / 0 \\
0.10 & 4 & 7 / 0 & 0 / 0 & 0 / 0 & 0 / 0$^\dagger$ \\
0.10 & 5 & --    & --    & --    & --    \\
\bottomrule
\end{tabular}\\[3pt]
\begin{minipage}{0.92\textwidth}\footnotesize
Confusion matrix over 500 cases (certify $=$ positive, closed-loop feasible $=$ actual positive): TP$=122$, FP$=30$, FN$=12$, TN$=336$ $\Rightarrow$ precision $0.80$, recall $0.91$, accuracy $0.92$, $F_1=0.85$, Matthews correlation $0.80$; false positives are $30/500\approx6\%$ of cases (false-positive rate $30/366\approx8\%$).\\
$^\dagger$Only 5 cases (not 9). The row $\omega_t = 5$~deg/s, $a_{\max} = 0.10$ is omitted: at this rate both the certificate and the closed-loop MPC collapse to near-empty (IoU $\le0.008$), so it carries no discriminating information.
\end{minipage}

\end{table*}

The prevalence of zero cells is not missing data; it is the predicted synchronization-radius collapse. A cell can be \emph{certified} only when the initial range lies inside the synchronization radius, $r_0 < r_{\mathrm{sync}} = 2a_{\max}/\omega_t^2$; beyond it the chaser cannot cancel the co-rotation velocity within the available range, and the certificate declines. The closed-loop column need not follow, because the empirical label asks only for $60$~s of LOS survival, which does not require co-rotation: the MPC can therefore still be feasible beyond $r_{\mathrm{sync}}$. The $0/8$ cell at $a_{\max}=0.02$, $\omega_t=1$~deg/s, $r_0=150$~m is exactly this case ($r_{\mathrm{sync}}\approx131$~m), and since none of its cases are certified it supplies $8$ of the $12$ false negatives on its own -- the certificate is conservative there, as a sound inner bound should be. For the swept points $r_{\mathrm{sync}}$ ranges from $657$~m ($a_{\max}=0.10$, $\omega_t=1$~deg/s) down to $5$~m ($a_{\max}=0.02$, $\omega_t=5$~deg/s); e.g.\ at $a_{\max}=0.02$, $\omega_t=2$~deg/s, $r_{\mathrm{sync}}\approx 33$~m, so only the $r_0=30$~m column is certified, and at $a_{\max}=0.10$, $\omega_t=3$~deg/s, $r_{\mathrm{sync}}\approx 73$~m, so only $r_0\in\{30,50\}$~m are certified, matching the table exactly. The non-empty cells therefore trace the feasible operating envelope, and the table is read row-wise as the envelope shrinking $\propto\omega_t^{-2}$.

Overall counts (out of 500 cases): 122 certified-safe, 12 empirical-only, and 366 unsafe. Treating ``certified'' as a binary predictor of closed-loop feasibility gives the confusion matrix $\mathrm{TP}=122$, $\mathrm{FP}=30$, $\mathrm{FN}=12$, $\mathrm{TN}=336$, hence precision~$0.80$, recall~$0.91$, accuracy~$0.92$, $F_1=0.85$, and Matthews correlation~$0.80$. The operationally relevant figure is that $\sim 6\%$ of all cases ($30/500$) are false positives (false-positive rate $\mathrm{FP}/(\mathrm{FP}+\mathrm{TN}) = 30/366 \approx 8\%$), concentrated at high tumble rate (e.g., $\omega_t = 4$~deg/s with $r_0 = 30$~m gives 7/9 certified but 0/9 feasible due to closed-loop transient violations during the co-rotation handover). The analytical criteria are therefore a useful, high-recall first-cut estimate, not a sufficient certificate.

\paragraph{Mechanism of the transient false positives.} The certified-but-infeasible cases (e.g.\ $\omega_t = 4$~deg/s, $r_0 = 30$~m: 7/9 certified, 0/9 flown) fail through two coupled controller-level effects that the open-loop analytical bound does not model. First, the deployed controller is the single-input safety filter~\eqref{eq:mpc_ocp}, which projects its command onto the thrust disk by scaling the \emph{entire} vector after the solve. During the close-approach handover the chaser must brake radially \emph{and} build along-track speed to hold the rotating port, so the filter commands both components at once; when their vector sum exceeds $a_{\max}$, the Euclidean projection scales the whole command down to $\|\mathbf{u}\|_2 = a_{\max}$, leaving the radial component below the value the open-loop bound assumed. The erosion test~\eqref{eq:erosion} credits the full disk authority $a_{\max}$ in the binding direction, whereas the realised command must share that budget with the orthogonal along-track demand, so the slack predicted positive by~\eqref{eq:erosion} is exhausted before the maneuver completes. Holding a single command across the forecast horizon, rather than planning a corrective input sequence, compounds the deficit.

Second, the discrete prediction enforces the LOS constraint only at the horizon nodes $t_k + j\Delta t$, while the corridor boundary rotates continuously at $\omega_t$. Between consecutive nodes the cone sweeps through $\omega_t \Delta t$ (at $\omega_t = 4$~deg/s, $\Delta t = 2$~s this is $8^\circ$), so a state that is feasible at both bracketing nodes can exit the corridor in the open interval, an inter-sample violation invisible to the QP. The two effects compound at small range, where the apparent angular rate of the boundary is largest and the diagonal-authority deficit bites hardest, which is precisely where the 6\% false positives concentrate. Both are properties of the realized discrete-time filter, not of the synchronization bound: a finer $\Delta t$, a genuine sequence-optimising MPC, or a continuous-time constraint-tightening margin recovers the certified cases at the cost of online compute.

\subsection{Hold-Mode Region of Attraction}\label{sec:roa}
For completeness, the unconstrained hold-mode LQR's region of attraction is the largest sublevel set $\{x : x^\top P x \le \rho^\star\}$ fitting inside the LOS cone, with $P$ the solution of the CWH discrete algebraic Riccati equation. This complements the safe-start analysis: once the chaser enters this set, the LQR keeps it there without further constraint management, providing a terminal certificate in the sense of~\cite{ReistTedrake2010,Maywald2021,Shala2024}. A detailed treatment is deferred to the planned journal extension.

\section{Discussion}

\subsection{Physical Honesty of the Simulation}
A significant finding during development was that earlier implementations using double-integrator truth dynamics with reference blending and state projection produced artificially successful results. In those versions, ``phantom'' delta-v from state teleportation accounted for up to 80\% of the apparent fuel expenditure, and scenarios that should have been dynamically infeasible (e.g., $\omega_t = 10$~deg/s, $a_{\max} = 0.02$~m/s$^2$, $r_{\mathrm{sync}} = 1.3$~m, starting from 195~m) reported successful hold. The transition to CWH truth dynamics with no blending or projection eliminated these artefacts and revealed the true physical limits of the guidance architecture.

\subsection{CWH Along-Track Coupling}
The dominant challenge for long-range approach is the CWH along-track coupling. The state-transition matrix contains the secular term $6(s - n\tau)$ in the $(2,1)$ element, which produces along-track drift proportional to the radial offset. At 195~m initial range, even a radial offset of a few meters produces substantial along-track acceleration that the controller must overcome. This effect is absent in double-integrator models, explaining why they produce overly optimistic results.

\subsection{Role of the Synchronization Radius}
The synchronization radius $r_{\mathrm{sync}} = 2a_{\max}/\omega_t^2$ emerges as the key design parameter governing feasibility. The benchmark of~\S\ref{sec:benchmark} confirms that the closed-form criteria capture the qualitative dependence on $\omega_t$ and $a_{\max}$ (shrinking IoU vs HJ as $\omega_t$ grows from 1 to 5~deg/s, mirroring the operational impossibility of co-rotation at high tumble rate). The precision~$0.80$ / recall~$0.91$ closed-loop agreement, together with the 250$\times$ speedup over HJ, supports the use of $r_{\mathrm{sync}}$ and the directional-erosion test as onboard go/no-go signals during mission design and replanning.

\paragraph{Inclusion of nominal, stochastic, and robust BRS variants.}
The didactic double-integrator example in \S\ref{sec:three_forms} establishes the
inclusion $\mathcal{X}_0^{\mathrm{robust}} \subseteq \mathcal{X}_0^{\mathrm{nom}}$,
with the stochastic set an empirical cloud bracketing the nominal boundary;
the robust variant is the operationally appropriate choice when
bounded thruster calibration error or navigation bias must be tolerated,
while the stochastic variant matches Gaussian sensor noise.
For this paper's headline benchmark we report the nominal BRS,
which is the largest of the three and therefore the most permissive
comparison against Hamilton--Jacobi.
For the CWH + rotating-LOS problem this conservative-by-construction ordering
specializes to the strict containment
\begin{equation}
\mathcal{S}_{\mathrm{tube}}\ \subseteq\ \mathcal{S}_{\mathrm{ana}}\ \subseteq\
\mathcal{S}_{\mathrm{sync}}\ \subseteq\ \mathcal{S}_{\mathrm{HJ}}^{\mathrm{pos}},
\label{eq:hierarchy}
\end{equation}
where $\mathcal{S}_{\mathrm{ana}}$ is the analytical synchronization set,
$\mathcal{S}_{\mathrm{tube}}$ its tube-tightened inner approximation (a robust
BRS with disturbance set $\mathcal{W}=$ steady-state mRPI box),
$\mathcal{S}_{\mathrm{sync}}$ the exact reach-and-co-rotate set, and
$\mathcal{S}_{\mathrm{HJ}}^{\mathrm{pos}}$ the position-only HJ backward reachable
set. The analytical criterion is thus a \emph{sound} (inner) bound on
synchronization, itself a strict subset of positional reachability.

\section{Planned Extensions}\label{sec:future}
A journal-length extension is planned that lifts the present results in four directions:
\begin{enumerate}
 \item \textbf{Y-A dynamics} for elliptical reference orbits, with the analytical synchronization radius extended to the Perron--Frobenius admissibility bound $\rho(\bar A(e)) < 1$, $e^\star \approx 0.661$, per~\cite{Iskender2026EllipticalTube,iskender2026horizon}.
 \item \textbf{Full 3D tumble} with general attitude kinematics.
 \item \textbf{Tube-MPC formulation} treating the rotating LOS corridor as a time-varying constraint and the truth-prediction mismatch as bounded disturbance.
 \item \textbf{Velocity-aware HJ target} adding a velocity-matching sub-constraint to the LOS cone target, closing the gap between the HJ ground truth and the analytical synchronization criteria.
\end{enumerate}

\section{Conclusions}
This paper established, for approach to a tumbling target under a rotating LOS corridor, a structural decoupling between two safe-start notions: the closed-form \emph{synchronization} set (per-constraint directional erosion~\eqref{eq:erosion} together with the synchronization range bound $r<2a_{\max}/\omega_t^2$~\eqref{eq:rsync}) is a sound inner certificate, strictly contained in the \emph{positional reachable} set computed by Hamilton--Jacobi, $\mathcal{S}_{\mathrm{sync}}\subsetneq\mathcal{S}_{\mathrm{reach}}$. The closed-form construction is $\sim 250\times$ faster than HJ; across a 500-case closed-loop sweep it predicts MPC feasibility with precision~$0.80$, recall~$0.91$, and accuracy~$0.92$, the residual 6\% false positives concentrating at high tumble rate, where the single-input filter's post-solve thrust projection and the horizon discretization induce transient corridor violations (\S\ref{sec:benchmark}). The analytical criteria therefore serve as an onboard go/no-go bound for active debris removal and on-orbit servicing, with HJ supplying the offline positional envelope. All results are reproducible from a single command.

\balance
\bibliographystyle{unsrt}
\bibliography{references}

\end{document}